\documentclass[aps,prx,showpacs,superscriptaddress]{revtex4-1}  
\usepackage{amssymb}
\usepackage[tbtags]{amsmath}
\usepackage{graphicx}
\begin{document}
\title{Topological quantization of the flow of magnetic skyrmions driven by a ratchet-like potential under thermal fluctuations }
\author{Shan-Chang Tang}
\affiliation{Department of Physics  \&  State Key Laboratory of Surface Physics, Fudan University,\\ Shanghai 200433, China}
\author{Yu Shi}
\affiliation{Department of Physics  \&  State Key Laboratory of Surface Physics, Fudan University,\\ Shanghai 200433, China}
\affiliation{Collaborative Innovation Center of Advanced Microstructures, Fudan University, \\Shanghai 200433, China}
\begin{abstract}
We consider a magnetic skyrmion adiabatically driven by a spin-polarized electrical current  periodic in both space and time and asymmetric in space, and also subject to a random magnetic field representing the thermal fluctuations. We show that when the random magnetic field is low enough,  while the  time variation of the driving current is slow enough,  the skyrmion flow is an integer multiply of the ratio between the space and time periods, the integer being a topological invariant called Chern number. This result is also demonstrated by numerically solving  the stochastic Landau-Lifshitz-Gilbert (sLLG)   and Langevin equations. Our work suggests a novel method of manipulating skyrmions with topological stability.
\end{abstract}
\maketitle

Magnetic skyrmions have attracted a lot of attention both experimentally and theoretically~\cite{NagaosaTokura-893}.  After their discovery in the chiral magnetic material MnSi~\cite{MuhlbauerBinz-78}, many studies have been conducted, for example, on  visibility~\cite{YuOnose-161,YuKanazawa-83,Heinzevon-246}, on motion driven by spin polarized electric current \cite{Jonietz-1589,YuKanazawa-231,SchulzRitz-1004,
EverschorGarst-216,IwasakiMochizuki-87},  on the transport in the presence of temperature gradient  \cite{KongZang-104,LinBatista-1534}, etc. The thermal ratchet motion under temperature gradient \cite{MochizukiYu-44} and the deterministic motion in a  ratchet potential, i.e. at zero temperature, were studied~\cite{ReichhardtRay-1386}.  In this Letter, we consider a skyrmion driven by a spin-polarized electric current periodic both spatially and temporarily, especially, it is also subject to a random magnetic field representing thermal fluctuations. Hence the motion of the skyrmion can be described as a kind of thermal ratchet in a uniform temperature~\cite{Reimann-1591}. We will show both analytically and numerically that in some parameter regime,  the average velocity of the skyrmion is topologically quantized as the Chern number multiplied by a basic unit, robust under thermal fluctuations.

The dynamics of the constituent spins of the skrymion are described by the two-dimensional sLLG equation~\cite{IwasakiMochizuki-87,KongZang-104,TroncosoNunez-300,
TroncosoNunez-1405,Garcia-1597},
\begin{equation}
\frac{\partial\mathbf{n}}{\partial t}+(\mathbf{h}\cdot\nabla)\mathbf{n}=-\frac{1}{\hbar}\mathbf{n}\times(\mathbf{H}_{eff}+\mathbf{R})+\alpha\mathbf{n}\times\frac{\partial\mathbf{n}}{\partial t}+\beta\mathbf{n}\times(\mathbf{h}\cdot\nabla)\mathbf{n}, \label{sllg}
\end{equation}
where $\mathbf{n}$ represents the spin direction of the spin,  with $|\mathbf{n}|^2=1$, $\alpha$ represents the damping effect, $\beta$ is the non-adiabatic coefficient,  $\mathbf{h}=-\frac{a^3}{2e}\mathbf{j}$, $a$ being the lattice constant and $\mathbf{j}$ being the spin-polarized current,  $\mathbf{H}_{eff}\equiv -\frac{\delta\mathcal{H_S}}{\delta\mathbf{n}}$, where  $\mathcal{H_S}=\iint dxdy\left[\frac{J}{2}(\nabla\mathbf{n})^2+
\frac{D_{DM}}{a}\mathbf{n}\cdot(\nabla\times\mathbf{n})-
\frac{1}{a^2}\mathbf{B}
\cdot\mathbf{n}-\frac{K}{a^2}n_z^2\right]$,  $J$ being the ferromagnetic coupling constant, $D$ being the Dzyaloshinski-Moriya interaction constant~\cite{Dzyaloshinsky-1598,Moriya-1599},   $K$ being the anisotropic constant, $\mathbf{B}$ being the external magnetic field, $\mathbf{R}$ is the random magnetic field, with $\langle R_i(\mathbf{r},t)\rangle=0$, $\langle R_i(\mathbf{r},t)R_j(\mathbf{r}',t')\rangle=2\alpha\hbar k_BTa^2\delta_{ij}\delta(\mathbf{r}-\mathbf{r}')\delta(t-t')$,
$i,j=x,y$, $\hbar$ is the Planck constant, $k_B$ is the Boltzman constant.

Following Thiele's method~\cite{Thiele-1600}, one can  obtain  the equation of motion of the skyrmion, that is,  Langevin equation~\cite{TroncosoNunez-300}
\begin{equation}
\alpha_d\left[\mathbf{\dot{Q}}-\frac{\beta}{\alpha}\mathbf{h}\right]
+\alpha_m\mathbf{\hat{z}}\times[\mathbf{\dot{Q}}-\mathbf{h}]
=\mathbf{\nu} (t),  \label{sl}
\end{equation}
where $\mathbf{Q}=(X,Y)$ is the position of the skyrmion as a whole,
$\alpha_d\equiv \alpha\iint dxdy\left(\frac{\partial\mathbf{n}}{\partial x}\right)^2$
and
$\alpha_m\equiv \iint dxdy\mathbf{n}\cdot\left(\frac{\partial\mathbf{n}}{\partial x}\times\frac{\partial\mathbf{n}}{\partial y}\right)$ are determined by by the spin texture of the skyrmion, $\mathbf{\nu}$ is   due  to the random magnetic field, satisfying $
\langle\nu_i(t)\rangle=0$ and $\langle\nu_i(t)\nu_j(t')\rangle
=2\frac{\alpha_dk_BTa^2}{\hbar}\delta_{ij}\delta(t-t')$, $i,j=x,y$.

Corresponding to the Langevin equation,   the  Fokker-Planck(FP) equation~\cite{Risken-1585}, describing   the probability density $\rho(\mathbf{r},t)$,  can be obtained  as
\begin{equation}
-\frac{\partial\rho(\mathbf{r},t)}{\partial t}=D\mathcal{O}\rho(\mathbf{r},t),  \label{sfp}
\end{equation}
where $D \equiv \frac{\alpha_d k_B Ta^2}{\hbar(\alpha_m^2+\alpha_d^2)}$,
$\mathcal{O}=-\nabla^2+\frac{\partial}{\partial x}(C_1 h_x+C_2 h_y) +\frac{\partial}{\partial y}(-C_2 h_x+C_1 h_y)$ is   the Fokker-Planck operator, with
$C_1\equiv \hbar\frac{\frac{\beta}{\alpha}\alpha_d^2+\alpha_m^2}{\alpha_d k_B Ta^2}$, $C_2\equiv \hbar\frac{(\frac{\beta}{\alpha}-1)\alpha_d\alpha_m}{\alpha_d k_B Ta^2}$. The probability current density $\mathbf{\mathcal{J}}=(\mathcal{J}_X, \mathcal{J}_Y)$ is given by $
\mathcal{J}_X=D\left[(C_1 h_x+C_2 h_y)-\frac{\partial}{\partial x}\right]\rho$, $
\mathcal{J}_Y=D\left[(-C_2 h_x+C_1 h_y) -\frac{\partial}{\partial y}\right]\rho.$
The average velocity of the skyrmion can be obtained as
\begin{equation}
\langle\dot{\mathbf{Q}}\rangle=\iint\mathbf{\mathcal{J}}dxdy
\end{equation}

The Fokker-Planck equation can be solved by transforming the Fokker-Planck operator to an Hermitian operator, under the condition~\cite{Risken-1585} $
\frac{\partial}{\partial y}(C_1 h_x+C_2 h_y)
=\frac{\partial}{\partial x}(-C_2 h_x+C_1 h_y). $
For simplicity,   suppose that the electric current is along $x$ direction such that  $h_y=0$. Then this condition  is  simplified as $
\frac{\partial h_x}{\partial y}+\kappa\frac{\partial h_x}{\partial x}=0,$
where $\kappa \equiv \frac{C_2}{C_1}$.
Hence $
h_x(x,y,t)=f(x-\kappa y)+g(t).$

For convenience, we switch to coordinates $(u,v)$ defined as
\begin{equation}
\begin{pmatrix}u\\v\end{pmatrix}=\frac{1}{\sqrt{1+\kappa^2}}\begin{pmatrix}1&-\kappa\\ \kappa&1\end{pmatrix}\begin{pmatrix}x\\y\end{pmatrix}. \label{transform}
\end{equation}
Then the Fokker-Planck operator can be written as
\begin{equation}
\mathcal{O}=-\nabla^2
+\xi\sqrt{1+\kappa^2}\frac{\partial}{\partial u}[f(\sqrt{1+\kappa^2}u)+g(t)],\label{fpo2}
\end{equation}
where $\xi=
\frac{1}{D}\frac{\frac{\beta}{\alpha}\alpha_d^2+\alpha_m^2}{
\alpha_d^2+\alpha_m^2}=
\frac{\hbar(\frac{\beta}{\alpha}\alpha_d^2+\alpha_m^2)}{
k_B T\alpha_d a^2}$, $\nabla^2=\frac{\partial^2}{\partial u^2}+\frac{\partial^2}{\partial v^2}$. The probability current density can be written as $\mathbf{\mathcal{J}}=(\mathcal{J}_u, \mathcal{J}_v)$, with
\begin{align}
\mathcal{J}_u&=D\left\{\xi\sqrt{1+\kappa^2}[f(\sqrt{1+\kappa^2}u)+g(t)]-\frac{\partial}{\partial u}\right\}\rho,\\ \label{Ju1}
\mathcal{J}_v&=-D\frac{\partial}{\partial v}\rho.
\end{align}
Clearly  $\langle\dot{v}\rangle=\iint\mathcal{J}_vdudv=0$.

In terms of
\begin{equation}
\Gamma(u,t)=-\xi\sqrt{1+\kappa^2}\int_0^udu'f(\sqrt{1+\kappa^2}u')-\xi\sqrt{1+\kappa^2}g(t)u, \end{equation}
the Fokker-Planck operator $\mathcal{O}$ can be transformed to a Hermitian operator
\begin{equation}
\mathcal{H}=e^{\Gamma/2} \mathcal{O} e^{-\Gamma/2}
=-\nabla^2+U,  \label{hmo}
\end{equation}
where $U=-\frac{1}{2}\frac{\partial^2\Gamma}{\partial u^2}+\frac{1}{4}\left(\frac{\partial\Gamma}{\partial u}\right)^2$. Consequently, the FP equation is transformed to a Schr\"{o}dinger-like equation
\begin{equation}
-\frac{\partial\psi(u,v,t)}{\partial t}=\left(D\mathcal{H}-\frac{1}{2}\frac{\partial\Gamma}{\partial t}-\frac{\partial\ln\sqrt{Z}}{\partial t}\right)\psi(u,v,t),  \label{sfp2}
\end{equation}
where $
\psi(u,v,t)=\rho(u,v,t)e^{\Gamma/2}\sqrt{Z}, $
with $Z=\iint e^{-\Gamma}dxdy$.

The  probability current on $u$ direction is
\begin{equation}
J_u(t)=\iint\mathcal{J}_u(u,v,t)dudv
=-2D\iint\psi_0(u,v,t)\frac{\partial}{\partial u}\psi(u,v,t)dudv. \label{probcrt}
\end{equation}

We now assume $U(u,v,t)=U(u+L,v,t)$, where $L$ is the space periodicity, and $U(u,v,t+\tau )=U(u,v,t)$, where $\tau$ is the time periodicity. We also assume $U$ to be asymmetric within each space period.  As such, our problem is  reduced to a two-dimensional generalization of a kind of thermal ratchet, that is, an  overdamped particle in an potential  periodic and asymmetric in space and periodic in time and also subject to a random force with thermal fluctuations, which is characterized by its nonzero unidirectional current~\cite{Reimann-1591}. For a one-dimensional thermal ratchet, it was demonstrated that in the so-called deterministic regime of the parameters and under the adiabatic condition,  the average velocity of the particle is quantized as a Chern number~\cite{ShiNiu-1596}, in analogy with the quantum adiabatic transport~\cite{Thouless-1601,niu1}. Under similar conditions, for the present problem of skyrmions, we now set out to show that its average velocity is also quantized as a Chern number. We also demonstrate this result numerically by using both sLLG and Langevin equations.

The instantaneous  eigenfunctions $\psi_{nk}(u,v,t)$ of  $\mathcal{H}$ satisfying
\begin{equation}
\mathcal{H}(u,v,t)\psi_{nk}(u,v,t)=E_{nk}(t)\psi_{nk}(u,v,t),
\end{equation}
where $E_{nk}(t)$ is the eigenvalue,  are Bloch waves
\begin{equation}
\psi_{nk }(u,v,t)=e^{ik  u}w_{nk }(u,v,t), \quad w_{nk }(u,v,t)=w_{nk }(u+L ,v,t), \end{equation}
where $-\frac{\pi}{L}\leqslant k <\frac{\pi}{L}$ is the magnitude of the Bloch vector. The ``ground state'' $\psi_0=\frac{1}{\sqrt{Z}}e^{-\Gamma(x,y,t)/2}$ represents the thermal equilibrium $\rho_0=\frac{1}{\sqrt{Z}}e^{-\Gamma(x,y,t)}$.

Under the adiabatic condition to be detailed later, the solution of the FP equation can be expanded in terms of these instantaneous eigenfunctions~\cite{Messiah-1602},
\begin{equation}
\psi(u,v,t)=\sum_{n,k }c_{nk }(t)
\psi_{nk }(u,v,t)e^{-D\int_0^tE_{0q }(t')dt'}.
\end{equation}
If the initial wave function is $\psi_{0q }$, then one obtains, up to the first order~\cite{Thouless-1601,ShiNiu-1596}, \begin{equation}
\psi_{q }(u,v,t)=\psi_{0q }(u,v,t)+\sum_{n\neq0,k \neq q }\frac{2\langle\psi_{nk }|\dot{\psi}_{0q }\rangle}{D(E_{0q }-E_{nk })}\psi_{nk },
\label{qq}
\end{equation}
where $\langle A|B\rangle \equiv \iint d^2\mathbf{r} A(u,v)^*B(u,v)$. Therefore the probability current is found to be
\begin{equation}
J_{q }(t)=\langle\partial_{q }w_{0q }|\partial_t w_{0q }\rangle-\langle\partial_t w_{0q }|\partial_{q }w_{0q }\rangle.
 \end{equation}

In fact,  the system is near the equilibrium, i.e. the ``ground state''. Thus  the physical probability current is given by $J_{0}(t)$.  Moreover, the probability density and thus the probability current adiabatically vary with time, following the variation of the  polarized electric current. Hence the   the probability current should   be averaged over the  time period,
\begin{equation}
\begin{split}
\overline{J_0}
&\equiv \frac{1}{\tau }   \int_0^{\tau } dtJ_0(t)
\\&=\frac{1}{\tau } \int_0^{\tau } dt[\langle\partial_{q }w_{0q }|\partial_t w_{0q }\rangle-\langle\partial_t w_{0q }|\partial_{q }w_{0q }\rangle]|_{q=0}
=-\frac{1}{\tau}\partial_q\theta (q)|_{q=0},
\end{split}
\end{equation}
where $\theta (q)= i\int dt \langle w_{0q}|\partial_t w_{0q}\rangle = i\int dt \langle \psi_{0q}|\partial_t \psi_{0q}\rangle$ is the Berry phase of $w_{0q}$ or $\psi_{0q}$ , under the assumption of adiabaticity.

In the deterministic limit,   $J_{q }(t)$ is  very insensitive to $q $, for the same reason that was given for the quantum adiabatic transport, by regarding $q$ as the parameter for the twisted boundary condition of $w_{0q}$, consequently $q$-dependent part of $J_{q }(t)$ is exponentially small~\cite{niu1}. Therefore $J_{0}(t)$ can be replaced as the average of $J_{q }(t)$ over different values of $q $,
\begin{equation}
\begin{split}
\overline{J_0}
&=\frac{1}{\tau }\frac{L}{2\pi}
\int_{-\frac{\pi}{L}}^{\frac{\pi}{L}}
dq \int_0^{\tau }dtJ_{q }(t)\\
&=\frac{L}{\tau }
\left\{\frac{1}{2\pi i}
\int_{-\frac{\pi}{L}}^{\frac{\pi}{L}}dq \int_0^{\tau } dt[\langle \partial_{q }w_{0q }|\partial_t w_{0q }\rangle-
\langle\partial_t w_{0q }|\partial_{q } w_{0q }\rangle]\right\}\\
&=\frac{L}{\tau }\mathcal{C}
\end{split}
\end{equation}
where $\mathcal{C}$ is the first Chern number, an integer. The last equality is valid because its a closed surface integral of a two-form, the Berry curvature.

Therefore, the average velocity of the skyrmion is given by  $\overline{\langle\dot{u}\rangle}=\frac{L}{\tau }\mathcal{C}$, $\overline{\langle\dot{v}\rangle}=0$,
that is,
\begin{equation}
\begin{split}
&
V_X=\frac{1}{\sqrt{1+\kappa^2} }\frac{L}{\tau }\mathcal{C}\\
&
V_Y=-\frac{\kappa}{\sqrt{1+\kappa^2} }\frac{L}{\tau }\mathcal{C}\label{chern}
\end{split}
\end{equation}
which indicates that the average velocity of the skrymion is quantized.

We now discuss the adiabatic condition that the rate $\frac{1}{\tau }$ of the time variation of the external potential should be much less than the spectrum gap $ D(E_1-E_0)$ between the ground state and the first excited state. Inspired by a previous simulation of a thermal rachet~\cite{Bartussek-1595}, we explicitly consider an example of the driving electric current,
\begin{equation}
j_x=\frac{2e}{a^2\tau_0}
\left\{-j_c\left[\cos G(x-\kappa y)+\frac{1}{2}\cos2G(x-\kappa y)\right]-A\cos\left(\frac{2\pi}{\tau }t\right)\right\},
\end{equation}
where $\tau_0 \equiv \frac{\hbar}{J}$ a unit of time, $\frac{2e}{a^2\tau_0}$ is a unit of the current density and thus $j_c$ is dimensionless.  $j_x$ is  periodic but asymmetric in $u$,  while $g(t)$ is periodic in time $t$, that is, $f(x-\kappa y)=\frac{a}{\tau_0}j_c
\left[\cos G(x-\kappa y)+\frac{1}{2}\cos2G(x-\kappa y)\right]$,  $g(t)=\frac{a}{\tau_0}A\cos(\frac{2\pi}{\tau }t)$.  Then it can be obtained that
\begin{equation}
\Gamma(u,t)
=-\frac{ \frac{\beta}{\alpha}\alpha_d^2+\alpha_m^2}{
\alpha_d \frac{k_B T}{J} a }
\left\{
\frac{j_c}{G}
\left[\sin G(\sqrt{1+\kappa^2}u)+\frac{1}{4}\sin2G(\sqrt{1+\kappa^2}u)
\right]
-\sqrt{1+\kappa^2} A\cos(\frac{2\pi}{\tau }t)u
\right\}.
\end{equation}
The order of magnitude   $\Gamma_0$ of the variation of $\Gamma$ can be estimated to be  $
\Gamma_0 \sim\frac{\frac{\beta}{\alpha}
\alpha_d^2+\alpha_m^2}{\alpha_d\frac{k_BT}{J}}\frac{L}{a}
\left(\frac{j_c}{2\pi}+A\right). $

When $\Gamma_0\ll 1$,  i.e. the temperature is high enough, the system is in the kinetic regime,  in which $E_1-E_0\sim\frac{1}{L^2}$, then the adiabatic condition becomes
\begin{equation*}
\frac{1}{\tau }\ll\frac{D}{L^2}=\frac{\alpha_d}{(\alpha_m^2+\alpha_d^2)}\frac{k_BT}{\hbar}
\left(\frac{L}{a}\right)^{-2}.
\end{equation*}
Thus the adiabatic condition in the kinetic regime is
\begin{equation}
\frac{\frac{\beta}{\alpha}
\alpha_d^2+\alpha_m^2}{\alpha_d} \frac{J}{k_BT} \left(\frac{L}{a}\right)
\left(\frac{j_c}{2\pi}+A\right) \ll 1,\,\,
\frac{\alpha_d^2+\alpha_m^2}{\alpha_d} \frac{J}{k_BT} \left(\frac{L}{a}\right)^2
\left(\frac{\tau_0}{\tau}\right) \ll 1.
\end{equation}

If we use typical parameter values  $\alpha_d\sim 1$, $\alpha_m\sim 10$, $L=100a$,
$j_c\sim 0.1$, $A\sim 0.1$, we have $
\Gamma_0\sim  10^3\left(\frac{k_BT}{J}\right)^{-1}$, the adiabatic condition in the kinetic regime becomes
$
\frac{k_BT}{J} \gg 10^3$ and $\frac{k_BT}{J} \gg
10^6  \left(\frac{\tau_0}{\tau}\right)$.

When $\Gamma_0\gg 1$,  i.e. the temperature is low  enough,  the system is in the  deterministic regime, in which  $E_1-E_0\sim\frac{\Gamma_0}{L^2}$.
Thus the adiabatic condition in the deterministic regime is
\begin{equation}
\frac{ \frac{\beta}{\alpha}
\alpha_d^2+\alpha_m^2}{\alpha_d} \frac{J}{k_BT} \left(\frac{L}{a}\right)
\left(\frac{j_c}{2\pi}+A \right) \gg 1,\,\, \frac{\frac{\beta}{\alpha} \alpha_d^2+\alpha_m^2}{
\alpha_d^2+\alpha_m^2} \left(\frac{L}{a}\right)^{-1}
\left(\frac{j_c}{2\pi}+A\right)  \left(\frac{\tau}{\tau_0}\right)\gg 1.
\end{equation}

It can be argued that in the deterministic regime, the nonadiabatic correction is exponentially small in $\Gamma_0$~\cite{ShiNiu-1596}. With the above parameter values,  the adiabatic condition in the deterministic regime is $
\frac{k_BT}{J} \ll 10^3$ and $\tau \gg 10^3\tau_0$.

In this regime,   we have  done  numerical simulations based on sLLG equation,  by using the lattice version of the Hamiltonian~\cite{IwasakiMochizuki-87,KongZang-104,NagaosaTokura-893}, $\mathcal{H_S}=-J \sum_{\langle ij\rangle}\mathbf{n}_i\cdot\mathbf{n}_j-D_{DM}\sum_{\langle ij\rangle}\mathbf{\hat{e}}_{ij}
\cdot{\mathbf{r}_i\times\mathbf{r}_j}-
\mathbf{B}\cdot\sum_i\mathbf{n}_i-K\sum_in_{iz}^2
$,
where $\langle ij\rangle$ represents the nearest-neighbour summation. We use the Runge-Kutta method on a $100\times 864$ lattice with the periodic boundary condition. The parameter values are the following, $G=\frac{2\pi}{100}$, $\kappa=\frac{100}{864}$, $\tau =5000\tau_0$,   $\alpha=0.1$, $B_z=0.015J$,   $D_{DM}=0.12J$, $K =0.01J$, $A=0.2$, $j_c$ is a tuning parameter. From the simulation, it is obtained that $\alpha_d\sim 1$, $\alpha_m\sim 10$, and thus the above condition $
\frac{k_BT}{J} \ll 10^3$ and $\tau \gg 10^3\tau_0$. applies.

\begin{figure}[h!]
  \centering
  \includegraphics[scale=0.3]{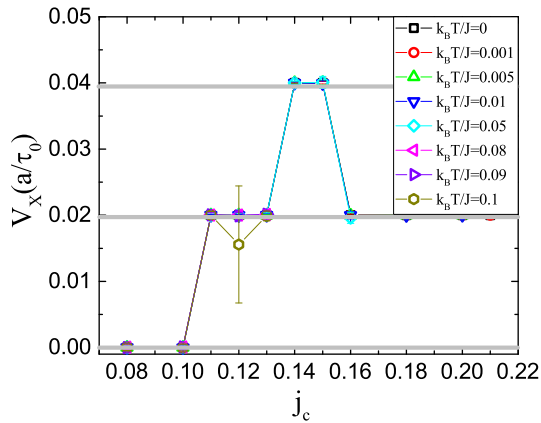} \\
  \caption{  Numerical simulation based on sLLG equation. Shown here is the dependence of the average  velocity along $x$  direction on the space-amplitude of the polarized electric current. The velocity is in the unit of $\frac{a}{\tau_0}$. Different symbols and colours represent different  values of $k_BT/J$. The grey lines represents the theoretically predicted values of the velocity. When $k_BT/J =0.1$,   the skyrmion itself becomes unstable. } \label{fig1}
\end{figure}

Our numerical result, as shown in Figure \ref{fig1},  clearly indicates that the average velocity of the skyrmion is quantized, exactly as our analytical result above has predicted. It can be seen that there is considerable deviation from the quantization at $k_BT/J=0.1$, since the skyrmion itself becomes unstable. Indeed, we have also done  numerical calculation by using  the Langevin equation, in which there is no such deviation.

To summarize, we have considered a skyrmion subject to a random magnetic field representing the thermal fluctuations, in addition to the usual forces and magnetic fields.  By designing the driving current to be ratchet-like, we have found that the average velocity of the skyrmion is quantized as first Chern number at low temperatures  such that the stochastic motion skyrmion is in the so-called deterministic limit and if the time variation of the driving electric current is slow enough. Our numerical calculations perfectly confirmed the quantized behaviour. This quantization provided a viable way to control the skyrmion at low temperatures, which could be useful for the the magnetic memory and communication. This work also combines the real space topological structure of spins with the momentum space topological structure of the motion. A very interesting aspect is that the adiabatic transport is shown to be a purely geometric effect, only depending on the ratio between the spatial and time periods, and on the topology of the instantaneous eigenfunctions of ${\cal H}$, while various parameters only determine the values to which $k_BT/J$ and $\tau/\tau_0$ are compared with, such that  the system is adiabatic in the deterministic regime.

This work is supported by National Natural Science Foundation of China (Grant Numbers 11374060 and 11574054).

\end{document}